\documentclass[prl,twocolumn]{revtex4}
\usepackage{graphicx}
\usepackage{amssymb,amsmath}

\begin{document}

\title{
Infinite matrix product states, Conformal Field Theory \\ and the Haldane-Shastry model}

\author{J. Ignacio Cirac $^{(1)}$ and
Germ\'an Sierra$^{(1,2)}$}

\affiliation{$^{(1)}$
Max-Planck-Institut f\"{u}r Quantenoptik, Hans-Kopfermannstr. 1, D-85748 Garching, Germany}
\affiliation{$^{(2)}$Instituto de F\'{\i}sica Te\'orica,
UAM-CSIC, Madrid, Spain}

\bigskip\bigskip\bigskip\bigskip

%
\font\numbers=cmss12
\font\upright=cmu10 scaled\magstep1
\def\stroke{\vrule height8pt width0.4pt depth-0.1pt}
\def\topfleck{\vrule height8pt width0.5pt depth-5.9pt}
\def\botfleck{\vrule height2pt width0.5pt depth0.1pt}
\def\Zmath{\vcenter{\hbox{\numbers\rlap{\rlap{Z}\kern
0.8pt\topfleck}\kern 2.2pt
                   \rlap Z\kern 6pt\botfleck\kern 1pt}}}
\def\Qmath{\vcenter{\hbox{\upright\rlap{\rlap{Q}\kern
                   3.8pt\stroke}\phantom{Q}}}}
\def\Nmath{\vcenter{\hbox{\upright\rlap{I}\kern 1.7pt N}}}
\def\Cmath{\vcenter{\hbox{\upright\rlap{\rlap{C}\kern
                   3.8pt\stroke}\phantom{C}}}}
\def\Rmath{\vcenter{\hbox{\upright\rlap{I}\kern 1.7pt R}}}
\def\Z{\ifmmode\Zmath\else$\Zmath$\fi}
\def\Q{\ifmmode\Qmath\else$\Qmath$\fi}
\def\N{\ifmmode\Nmath\else$\Nmath$\fi}
\def\C{\ifmmode\Cmath\else$\Cmath$\fi}
\def\R{\ifmmode\Rmath\else$\Rmath$\fi}
\def\H{{\cal H}}
\def\NN{{\cal N}}
\def\tv{{\tilde{v}}}
\def\vep{{\tilde{\epsilon}}}
\def\te{{\tilde{\vep}}}
\def\sh{{\rm sh}}
\def\cth{{\rm cth}}
\def\th{{\rm th}}
\def\bk{{\bf k}}
\def\br{{\bf r}}

\begin{abstract}
We generalize the Matrix Product States method using the chiral vertex operators of Conformal Field
Theory and apply it to study the ground states of the XXZ spin chain, the $J_1-J_2$ model and
random Heisenberg models. We compute the overlap with the exact wave functions, spin-spin  correlators
and the Renyi entropy, showing that critical systems can be described by this method.
For rotational invariant ansatzs we construct an inhomogenous
extension of  the Haldane-Shastry model with long range exchange interactions.
\end{abstract}


\maketitle

\vskip 0.2cm

In recent years the study of one-dimensional quantum lattice systems
has received an enormous impetus due the invention of novel
numerical algorithms, like DMRG \cite{DMRG} and its extensions
(for a review see \cite{cv09}).
The key for the success of those methods is the careful treatment
of the complex structure of the low energy states, which is based on the
behavior of the entanglement entropy at zero temperature \cite{ecp08}.
For a gapped system, the entanglement entropy, $S_L$,
of a subsystem of length $L$ converges towards a constant value,
independent of $L$ \cite{hastings}. For critical systems, however,
$S_L$ grows as the logarithm of $L$, were the
proportionality factor is related to the central charge $c$ of the underlying
Conformal Field  Theory (CFT) \cite{c}. This behavior of $S_L$
implies \cite{verstfaith} that the ground state of the system is well approximated by
a matrix product state (MPS) \cite{MPS}, a state characterized
in terms of certain matrices $A_s$. This explains the enormous success
of DMRG, since it can be understood as a variational method with respect
to MPS of bounded dimension \cite{RommerOstlund}. However, the
unbounded increase of the entanglement entropy makes DMRG much less accurate
for critical systems  since the size of the matrices characterizing the MPS
must inevitably grow with the size of the system. Thus, in order to properly
describe the low energy states of such systems MPS with infinitely dimensional matrices
$A_s$ are required.

The purpose of this letter is to construct such infinite dimensional MPS
(IMPS for short) and their application to 1D spin chains. We will replace
the finite dimensional matrices $A_s$ by chiral vertex operators of a CFT.
These operators act on infinite dimensional Hilbert spaces, which acquire the
meaning of ancillary space where entanglement and correlations are transported.
The huge enlargement of the ancillary space will
allow us to describe both critical and non-critical 1D systems
on equal footing. In particular, we will consider the XXZ and J1-J2 models
to illustrate the accuracy of our method.

Whereas standard MPS have, in general, a complicated form
if written in the spin basis, the IMPS that we shall consider below have a rather
simple form given by Jastrow type wave functions. There exists many important examples
of that type wave functions, such as the Laughlin
of the FQHE \cite{l83}, the Haldane-Shastry of the inverse square
Heisenberg model \cite{hs} and the related Calogero-Sutherland state of hard core bosons \cite{cs}.
A particular example of IMPS, which has spin rotational invariance,
will coincide with the Haldane-Shastry (HS) wave function. This case is particularly
interesting because there is an enhanced symmetry described by the $SU(2)$
Wess-Zumino-Witten model at level 1. Using the Ward identities of this CFT
we shall find an integrable extension of the HS Hamiltonian
parameterized by a large number of coefficients. We will then use this
extended HS model to study the entanglement entropy of spin chains with
frustration and random couplings, finding a very good agreement with a
law recently proposed in \cite{rm09,calabrese}.

Let us consider a
spin-$1/2$  chain with $N$ sites, and  local spin basis $|s_i \rangle$,
where $s_i= \pm 1 \; (i=1, \dots, N)$.   A generic state in the Hilbert space
of this spin chain is given by
\begin{eqnarray}
|\psi \rangle = \sum_{s_1, \dots, s_N} \psi(s_1, \dots, s_N) \; |s_1, \dots, s_N \rangle
\label{psi}
\end{eqnarray}
A MPS state is an ansatz of the form
\begin{eqnarray}
\psi(s_1, \dots, s_N) = \langle u |A^{(1)}_{s_1} \dots A^{(N)}_{s_N} |v \rangle
\label{mps}
\end{eqnarray}
where $A^{(i)}_{s_i}$  and $\langle u| , |v \rangle$ are $D$-dimensional matrices and row and column
vectors, respectively. The number or parameters needed to specify a MPS scales as $2 D^2 N$,  which
is much smaller than the number of components of a generic state (\ref{psi}), namely $2^N$.
To define a IMPS we shall replace the matrices $A^{(i)}_{s_i}$ by the chiral vertex operators
of a CFT, thus effectively working with $D=\infty$.
Let us choose for simplicity a gaussian CFT with central charge $c=1$.  Chiral vertex operators
are normal ordered exponentials of the chiral bosonic field $\varphi(z)$ \cite{cft-book},
which we can be used to define
\begin{eqnarray}
 A^{(i)}_{s_i} = \chi_{s_i}  : e^{ i s_i \sqrt{\alpha} \varphi(z_i)} : \; \;(i=1, \dots,N)
 \label{imps}
\end{eqnarray}
where $z_i$ are complex numbers, $\alpha$ a positive real number and $\chi_{s_i}= 1, s_i$ for
$i$ even and  odd respectively  \cite{marshall}. For the vectors
$\langle u| , |v \rangle$, we take the outgoing and incoming vacuum states
of the gaussian CFT. Equation (\ref{mps}) becomes then the vacuum expectation value
of a product of vertex operators which is given by
\begin{eqnarray}
\psi(s_1, \dots, s_N) = \delta_s e^{ i \pi/2 \sum_{i: {\rm odd} } ( s_i -1)}
\; \prod_{i > j}^N  (z_i - z_j)^{ \alpha s_i s_j}
\label{wave1}
\end{eqnarray}
where $\delta_s=1$ if $ \sum_{i=1}^N s_i =0$ and zero otherwise.
This condition can be traced back to
the conservation of the $U(1)$ charge of the gaussian CFT.
Equation(\ref{imps}) associates
the charge of the vertex operator $A^{(i)}_{s_i}$ to the local spin $s_i$, so that
the total third component of the spin, $S^z= \sum_i s_i/2$,  vanishes.
We will use the IMPS variationally, with $z_i$ and $\alpha$ as variational parameters.
The wave function  (\ref{wave1}) scales with an overall factor under a Moebius transformation
$ z_i \rightarrow (a  z_i+ b)/(c z_i + d) \; \forall i $, so that the ansatz only depends
on $N-3$ $z$-parameters. The latter parameters
must not be identified with the position of the local spins $s_i$.  An exception
is the wave function associated to translationally invariant spin chains where
we shall choose $z_n = e^{ 2 \pi i n/N} \; (n=1, \dots, N)$. In that case we have
\begin{eqnarray}
\psi(s_1, \dots, s_N)  \propto   \delta_s    e^{ i \frac{\pi}{2}  \sum_{i: {\rm odd} } s_i }   \; \prod_{n > m}^N  \left(
\sin \frac{ \pi (n-m)}{N} \right)^{ \alpha s_n s_m}
\label{wave2}
\end{eqnarray}
For later purposes,  it is convenient to express this
spin wave function using hard-core boson variables. Mapping the spin up (down)
states into empty (occupied) hard-core boson states, one finds
\begin{eqnarray}
\psi(n_1, \dots, n_{N/2})  \propto     e^{ i \pi   \sum_i n_i    }   \; \prod_{n_i > n_j}^{N/2}   \left(
\sin \frac{ \pi (n_i-n_j)}{N} \right)^{4  \alpha}
\label{wave3}
\end{eqnarray}
where  $n_i=1,\ldots,N$ denote the positions of $N/2$ hard-core bosons in the lattice.
\begin{figure}[t!]
\begin{center}
\includegraphics[width=.9\linewidth]{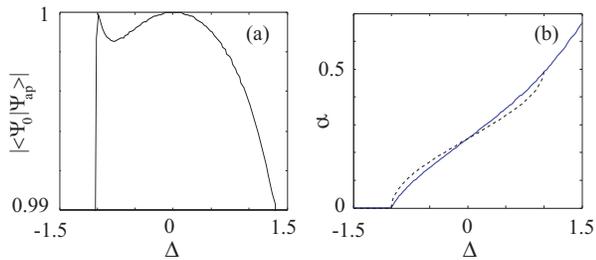}
\end{center}
\caption{
(a) Overlap of the variational ansatz (\ref{wave2}) with the exact GS wave function of the Hamiltonian
(\ref{XXZ}) for $N=20$ spins; (b) Optimal value of $\alpha$. The dashed line shows the curve
$\Delta=-\cos(2\pi\alpha)$.}
\label{overlap-XXZ}
\end{figure}

{\em 1.- The XXZ spin chain:} The  Hamiltonian with anisotropic coupling $\Delta$ is given  by
\begin{eqnarray}
H_{XXZ} = \sum_{i=1}^N \left(  S^x_i S^x_{i+1} +   S^y_i S^y_{i+1}+ \Delta  S^z_i S^z_{i+1}\right)
\label{XXZ}
\end{eqnarray}
where ${\bf S}_{N+1} = {\bf S}_1$ for  periodic boundary conditions (we take $N$ to be even).
This model displays three different phases at zero temperature: i) a gapped AF phase ($\Delta > 1$), ii) a critical phase
($-1 < \Delta \leq 1$) and iii) a ferromagnetic phase $\Delta \leq - 1$.
To describe the GS of the XXZ chain  we shall use the translational invariant state (\ref{wave2}),
which  favours antiparallel nearest-neighbour spins. We expect
this ansatz to adequately describe the $\Delta > -1$ regimes.
The only variational parameter  left in (\ref{wave2}) is $\alpha$,
which we fix by numerically minimizing the energy.
In Fig. \ref{overlap-XXZ} we plot this overlap as a function of $\Delta$,  for $N=20$.
As anticipated above, the overlap  is rather poor in the ferromagnetic regime but
surprisingly  good for $\Delta > -1$, even above $99 \%$.
Moreover, the overlap is exactly one at $\Delta = 0$ and $-1$, where $\alpha=0$ and $1/4$ respectively.
This indicates that the corresponding IMPS states coincide with the exact GS.
To prove this result, one applies the unitary transformation $U=\prod_{i:{\rm odd}}(2S_i^z)$ to (\ref{XXZ})
which flips the signs of the XY exchange interactions.
For $\Delta = -1$, one obtains  the isotropic ferromagnetic Heisenberg Hamiltonian,
whose GS is fully polarized with $S^z = N/2$. Applying the
lowering operator $(S^-)^{N/2}$ to this state,  and undoing the unitary
transformation,  one recovers  (\ref{wave2}) for $\alpha =0$.
In the case $\Delta =0$, the transformed Hamiltonian describes
free hard-core bosons, whose GS is the absolute value of a Slater determinant
which yields
\begin{eqnarray}
f(n_1, \dots, n_{N/2})
 \propto \prod_{n_i >n_j}^{N/2}    \sin \frac{ \pi (n_i - n_j)}{ N}
\label{hard}
\end{eqnarray}
where $n_1, \dots, n_{N/2}$ are the positions of the bosons on the lattice.
Undoing the $U$ transformation, one recovers the state (\ref{wave3}),
with $\alpha = 1/4$.
For  the isotropic XXX model, $\Delta =1$, the maximal overlap is obtained for $\alpha=1/2$.
The corresponding wave function (\ref{wave2}) does not yield the exact GS,  but
its hard-core version (\ref{wave3}) coincides with the GS of the Haldane-Shastry model,
which belongs to the same universality class that the XXX chain
(see below for a more detailed discussion).

\begin{figure}[t!]
\begin{center}
\includegraphics[width=.9\linewidth]{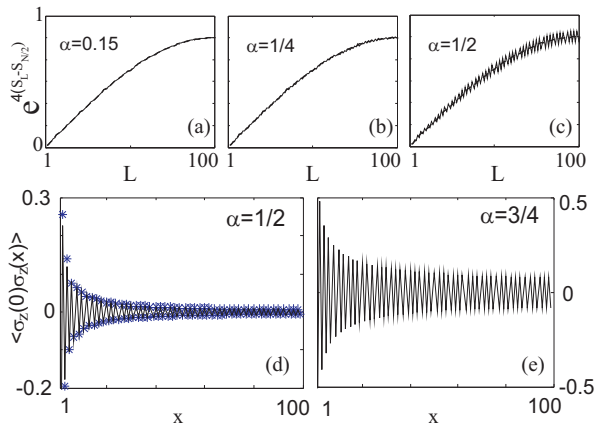}
\end{center}
\caption{(a,b,c)
Renyi entropy $S^{(2)}_L$ for a chain with $N=200$ sites and several values of
$\alpha \in (0, 1/2]$.
The dashed line shows the CFT prediction for $c=1$. (d,e) Correlator
$C^{zz}_n $ for several values of $\alpha$. In (d) the stars indicate
the exact result.}
\label{EntropyN200}
\end{figure}

Collecting the previous results we see that  the IMP ansatz (\ref{wave2}), in the range
$0 < \alpha \leq 1/2$, corresponds to the critical XXZ chain.
To confirm the critical properties of this ansatz  we have computed the Renyi entropy
$S^{(2)}_L = - \log {\rm Tr} \;  \rho^2_L$, where $\rho_L$ is the density matrix of
(\ref{wave2}) restricted to a subsystem of size $L$.
The computation is performed as follows. We rewrite
 \begin{equation}
 e^{-S^{(2)}_L} = \sum_{n,n',m,m'}
 |\langle n,m|\psi\rangle|^2 |\langle n',m'|\psi\rangle|^2
 \frac{ \langle\psi|n',m\rangle \langle\psi|n,m'\rangle}
 {\langle\psi|n,m\rangle \langle\psi|n',m'\rangle}
 \end{equation}
where $|n\rangle$ (and $|n'\rangle$) is an orthonormal basis in the space of the $L$ spins, and
$|m\rangle$ (and $|m'\rangle$) another corresponding to the rest of spins. This expression can be easily
evaluated by using two independent spin chains and performing the additions
using MonteCarlo techniques. In Fig. \ref{EntropyN200}(a,b,c)
we plot ${\rm exp}[ {4 (S^{(2)}_L - S^{(2)}_{N/2})/c}]$ for $N=200$ and several values of
$\alpha$. For a CFT with
central charge $c=1$ one expects this quantity to be $ \sin( \pi L/N)$ \cite{c},
which is also plotted (dashed line). The numerical results agree (in average) very
well with this prediction, confirming the criticality of the IMPS
for $\alpha\le 1/2$. Note that the oscillations are not a feature of the
numerical calculations but they are intrinsic in the model.
For $\alpha > 1/2$, we have checked that the Renyi entropy
saturates to a constant value independent of $L$, in accordance with the gapped
character of the XXZ spin chain with $\Delta > 1$.

Another quantities that can be easily computed using MonteCarlo techniques are
spin-spin correlators  $C^{aa}_n  = \langle \sigma^a_n \sigma^a_0 \rangle \; (a=x,y,z)$.
Fig.\ref{EntropyN200}(d,e) shows the correlator $C^{zz}_n $
for $\alpha =1/2, 3/4$.   In the case  $\alpha = 3/4$,
the correlator exhibits anti-ferromagnetic long range order, as expected for the gapped $\Delta > 1$
regime. For $\alpha=1/2$ we can compare with the exact result (stars in the figure)
$C^{zz}_n =  (-1)^n {\rm Si}( \pi n)/(\pi n)$
where ${\rm Si}(x)$ is the sine integral function. For this value of $\alpha$, the system
is isotropic (see below), so $C^{aa}_n$ is independent of $a$, and its expression
was first obtained from a Gutwiller proyection of the one-band
Fermi state \cite{gv87}. This result motivated  Haldane and Shastry to
construct their inverse square exchange Hamiltonian \cite{hs}. These authors also  noticed  that $C^{xx}_n$
coincides with the  one-body density matrix of the Calogero-Sutherland model of a gas of hard core bosons \cite{cs}.
The GS  of the latter model is a continuum version of (\ref{wave3}) for $\alpha = 1/2$.
For other values of $\alpha \in (0, 1/2]$, the asymptotic behaviour of  the one-body density matrix correlator
is $C^{xx}_n \sim n^{- 2 \alpha}$ \cite{agls06} , which can be compared with the
exact scaling in the critical XXZ model $C^{xx}_n \sim n^{- \eta} \; ( \Delta = - \cos{ \pi \eta})$ \cite{lp75}.
The comparison of these wo results
yields,   $\Delta = - \cos( 2 \pi \alpha)$, which correctly
reproduces the cases: $\alpha=0,  1/4,1/2$ (see Fig. \ref{overlap-XXZ}(b)).

{\em 2.- $J_1-J_2$ Heisenberg model:} We use the IMPS ansatz for this frustated
antiferromagnet model described by the Hamiltonian
 \begin{eqnarray}
 H_{J_1, J_2}  =  \sum_{i=1}^N  ( J_1   {\bf S}_i \cdot {\bf S}_{i+1} +  J_2   {\bf S}_i \cdot {\bf S}_{i+2} )
  \label{j1j2}
 \end{eqnarray}
\begin{figure}[h!]
\begin{center}
\includegraphics[width=.9\linewidth]{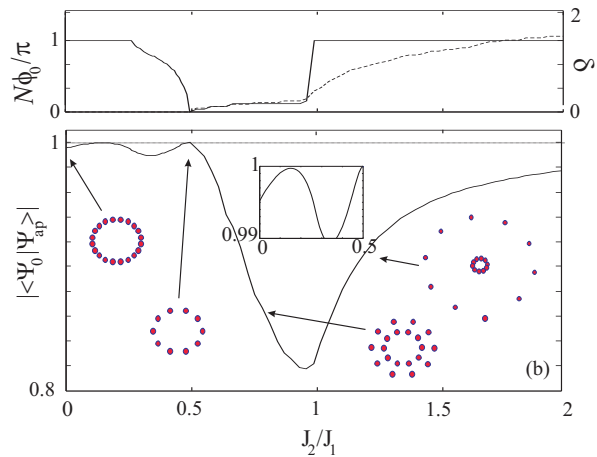}
\end{center}
\caption{$J_1-J_2$ model (a) Values of $\phi_0$ and $\delta$ for which the energy is minimal;
(b) Overlap of the IMPS state with the exact GS. The values of the $z_n$ in the
complex plane are qualitatively depicted for each region.}
\label{delta-phi}
\end{figure}

The phases of this model are: i) critical $c=1$ phase $(0 \leq J_2 \leq J_{2,c} \sim 0.241$,
ii) spontaneously dimerized phase ($J_{2,c} < J_2 \leq J_{\rm MG} = 0.5)$ and  iii) incommesurate spiral phase
($J_2 > J_{MG}$) \cite{wa96}. $J_{\rm MG}$  is the Majumdar-Gosh point whose two GS are the dimer configurations.
Since this model is isotropic we shall choose the ansatz (\ref{wave1}) with  $\alpha = 1/2$
(see \cite{fs97} for a related anstaz). For the choice of the $z_i$ parameters we distinguish between even and odd sites
 \begin{eqnarray*}
z_{2 n} = e^{ \delta - i \phi_0 +  \frac{ 4 \pi i n}{N}}, \;
z_{2 n+1} = e^{ -\delta + i \phi_0 +  \frac{ 4 \pi i n}{N}}, \;n=1, \dots, \frac{N}{2}
  \label{zn}
 \end{eqnarray*}
where $\phi_0 \in [0, \pi/N]$ and $\delta$
are variational parameters, which for a translational invariant state are
$(\phi_0, \delta) = (\pi/N, 0)$. These parameters are found minimizing the
energy, and their value (a) as well as the overlap with the exact GS wave function (b)
are shown in Fig.\ref{delta-phi} as functions of  $J_2/J_1$.
At  $J_2 \sim J_{2,c}$, the parameter $\phi_0$
departures from $\pi/N$, which reflects the spontaneous dimerization of the system.
At the MG point,  $\phi_0=\delta=0$ and the IMPS is a linear combination
of the dimer states. For $J_2 > J_{MG} $ the parameter $\delta$ departures
from zero. Finally,  at $J_2/J_1 \sim 1$, one finds $\delta= \pi/N$,
while $\delta$  increases steadly with an overlap approaching one, meaning that
the two chains become increasingly decoupled. A pictorial representation of the
variational parameters is also shown in fig.\ref{delta-phi}(b).

{\em 3.- Inhomogeneous Haldane-Shastry Model:}
Each (finite)  MPS is the GS of a finite range parent (frustration free) Hamiltonian \cite{cv09}.
Hence one may expect the IMPS parent Hamiltonians  to  have a range comparable to the size
of the system. We will show that this is the case for the case $\alpha =1/2$ and
arbitrary $z_n$. Remarkably, the corresponding parent Hamiltonian has two--spin interactions
only and is a generalization of the HS model.
First notice that the wave function (\ref{wave1})
is the chiral conformal block of  $N$ primary fields, of spin 1/2 and conformal weight  $h=1/4$,
in the $SU(2)_1$ WZW model.
These primary fields are nothing but the chiral vertex operators (\ref{imps}) for $\alpha = 1/2$.
The chiral conformal blocks of the WZW model, on a cylinder,  satisfy the Knizhnik-Zamolodchikov equations,
 \begin{eqnarray}
\left(  \frac{k+2}{2} z_i \frac{ \partial}{\partial z_i} -
   \sum_{j \neq i} \frac{z_i + z_j }{ z_i - z_j}   {\bf S}_i \cdot {\bf S}_j   \right)\; \psi_{\rm cyl}( {\bf z}) =0
  \label{kz}
 \end{eqnarray}
where $\psi_{\rm cyl}( {\bf z}) = \prod_i^N z_i^{1/4}  \psi( {\bf z})$ is a conformal transformation
of the wave function (\ref{wave1}) from the complex plane into the cylinder.
Now, using eqs (\ref{kz}) and (\ref{wave1}) one can prove that   $\psi_{\rm cyl}( {\bf z})$
is an eigenstate of the  Hamiltonian
 \begin{eqnarray}
H = - \sum_{i \neq j} \left(  \frac{ z_i z_j}{ (z_i - z_j)^2} + \frac{ w_{i j} ( c_i - c_j)}{ 12} \right)
  {\bf S}_i \cdot {\bf S}_j
  \label{parent}
 \end{eqnarray}
 with eigenenergy
 \begin{eqnarray}
E =  \frac{1}{16}  \sum_{i \neq j} w_{i j}^2 - \frac{N ( N+1)}{16}
  \label{energy}
 \end{eqnarray}
where $w_{ij} = (z_i + z_j)/(z_i - z_j), c_i = \sum_{j \neq i} w_{ij}$.
Taking $z_n = e^{ 2 \pi n/N}$, the parameters $c_i$ vanish and (\ref{parent})
becomes the HS Hamiltonian whose GS is indeed (\ref{wave2}).  For other choices of $z_n$, we have checked
numerically that  $\psi_{\rm cyl}$ is the GS of (\ref{parent}). The Hamiltonian (\ref{parent})
is an inhomogeneous generalization of the HS model, which seems to be integrable due to the
existence of higher conserved quantities that can be derived from the KZ equation (\ref{kz}).
The uniform HS model has a huge degeneracy in the spectrum that can be explained by a
Yangian symmetry. This degeneracy is broken in the non uniform case.

\begin{figure}[h!]
\begin{center}
\includegraphics[width=.9\linewidth]{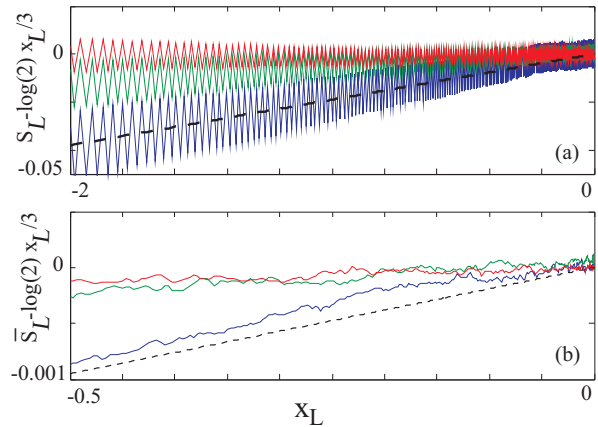}
\end{center}
\caption{Renyi entropy $S^{(2)}_L$ as a function of $x_L=\log[\sin(\pi L/N)]$. We
have substracted from the entropy $\log(2) x_L/3$ in order to compare with the prediction
\cite{rm09,calabrese}. Upper (red), medium(green), and lower (blue) curves correspond to
$\delta=0.75,0.5,0.1$. The dashed line corresponds to $x_L/4$. In (b) we have averaged
over consecutive points and plotted for low values of $x_L$.}
\label{Randomentropy}
\end{figure}

We can use the above construction to study the entanglement entropy of random models.
The scaling of the von Neumann entropy for the random AFH model is $ \frac{1}{3}  \log 2  \log L$ 
 \cite{rm04}, and it is conjectured that the same law holds for the Renyi entropy  \cite{rm09,calabrese}.
We take the IMPS ansatz
(\ref{wave1}) with $\alpha = 1/2$ and $z_n = e^{2 \pi i (n + \phi_n)/N}$,
and random choices of $\phi_n$ uniformly distributed in $[-\delta/2,\delta/2]$.
This corresponds to having the inhomogeneous HS model with random couplings.
We have computed $S^{(2)}_L$ averaged over realizations and plotted it in
Fig. \ref{Randomentropy} for $N=1000$. We have subtracted $ \frac{1}{3}  \log 2  \log L$ to
compare with the prediction \cite{rm09,calabrese} which should yield a horizontal line
at 0. For $\delta=0.75, 0.5$, if we ignore the oscillations, we find a very
good agreement. Even though for $\delta=0.1$ the result seems to fit better
with the formula $ \frac{1}{4}  \log L$, the inset shows that for long distances
the previous result seems to apply. Thus, the larger the $\delta$ the more valid
the prediction is for shorter blocks.

In this letter we have proposed an infinite dimensional version of the MPS using the chiral  vertex operators
of a $c=1$ CFT. This generalization allow us to study the entanglement properties and correlators of critical
and non critical spin chains. For isotropic spin chains the CFT is the $SU(2)_1$ WZW model
and this fact allow us to construct their associated parent Hamiltonians. They 
are given by an inhomogenous generalization of the Haldane-Shastry
Hamiltonian, that seems to be integrable. Our construction can be  generalized in principle
to any CFT. For rational CFTs, such as the WZW models, we expect that the conformal
blocks may provide degenerate GS with non-abelian statistics encoded in their braiding matrices.

{\it Acknowledgments-}
We thank  P. Calabrese and M. Hastings, for useful discussions.
This work was supported by the projects FIS2006-04885, EU STREP (Quevadis),
 DFG (FOR635), and by the ESF INSTANS 2005-2010.


\begin{thebibliography}{999}
\bibitem{marshall} This choice is called the Marshall sign rule which holds in the
antiferromagnetic XXX model.
\end{thebibliography}
\end{document}